\documentclass[conference]{IEEEtran}

  	\usepackage[pdftex]{graphicx}
  	\graphicspath{{../pdf/}{../jpeg/}}
	\DeclareGraphicsExtensions{.pdf,.jpeg,.png}
	\usepackage{algorithm}
	\usepackage{algpseudocode}
	\usepackage[cmex10]{amsmath}
	\usepackage{array}
	\usepackage{mdwmath}
	\usepackage{mdwtab}
	\usepackage{eqparbox}
	\usepackage{url}
\begin{document}
\title{\LARGE Analyzing Bridge Resonance and Lateral Vibrations Using String Vibration Principles}

 \author{\authorblockN{Zhang Jianan\authorrefmark{1}, Wang Yiyi\authorrefmark{2}, Duan Hongyi\authorrefmark{2},Li Yuchen\authorrefmark{2} }
 \authorblockA{\authorrefmark{1}School of Mathematics, Shanghai University of Finance and Economics, Shanghai 200433, China}
 \authorblockA{\authorrefmark{2}Faculty of Electronic and Information, Xi'an Jiaotong University, Xi'an, 710049, China\\ziaqifei@ieee.org}}

\maketitle

\begin{abstract}

This article investigates the lateral vibration and resonance of bridges, crucial for transportation network integrity and traffic safety. It aims to understand the underlying principles and causes of these vibrations to enhance bridge design and maintenance. Utilizing Euler-Bernoulli beam theory and assumptions like linear elasticity and uniform material, the study simplifies complex bridge dynamics into a manageable model. It explores the bridge's flexural response under static and dynamic loads, focusing on resonance phenomena.

Numerical simulations, including the finite difference method and analysis of nonlinear elastic responses, assess the bridge behavior under various load conditions, particularly periodic loads. The findings offer theoretical insights and practical guidelines for vibration control and safe bridge operation.
\end{abstract}

\IEEEoverridecommandlockouts

\IEEEpeerreviewmaketitle

\section{Introduction}

Bridges, as pivotal components of contemporary transportation infrastructure, not only bear significant traffic loads but also confront a multitude of challenges stemming from environmental factors and external forces. Among these, lateral vibration emerges as a paramount concern in the realm of bridge engineering.\cite{chen2020multimode} This phenomenon, manifesting in bridges, leads to structural fatigue and potential damage, while adversely impacting vehicular safety and the efficiency of traffic flow. Consequently, a thorough investigation into the dynamics and causative factors of bridge lateral vibrations is imperative for ensuring their structural integrity and operational reliability.\cite{wei2022lateral}

Within the domain of vibration mechanics, the theory of string vibration has long been established as a classical model for examining resonance and vibrational issues. The string vibration model's strength lies in its capacity to precisely delineate structural vibratory behavior through mathematical methodologies, encapsulating critical parameters such as frequency, amplitude, and mode of vibration. Nevertheless, in practical bridge engineering applications, the lateral vibration conundrum is exacerbated by the complexity and heterogeneity of bridge architectures. Influential factors include the geometric configuration, elastic properties, support conditions, and the external loading borne by the bridge.\cite{fujino2019research} Hence, the analysis of lateral vibrations in various bridge types necessitates the development of bespoke mathematical models, complete with tailored initial and boundary conditions.

The primary objective of this article is to investigate the patterns of lateral vibration in bridges and to demystify the underlying causes of resonance through the establishment of appropriate models. This exploration will concentrate on the vibrational characteristics of bridges, encompassing variations in frequency and amplitude, as well as the analysis of transverse vibration modes. Specifically, this study will account for the presence of elastic bearings at both extremities of the bridge, a prevalent support mechanism in practical engineering scenarios. Moreover, the article will evaluate the influence of external forces on bridge vibrations to acquire a holistic understanding of the mechanisms governing bridge lateral vibrations.

By delving into the lateral vibration phenomenon in bridges, this research aspires to furnish engineers and scholars with critical insights into bridge vibratory characteristics, thereby facilitating the enhancement of bridge design and maintenance methodologies. Furthermore, elucidating the origins of resonance will augment our comprehension and management of lateral vibration issues, contributing significantly to the sustainable progression of bridge engineering.

In essence, this investigation seeks to elucidate the fundamental principles underlying bridge lateral vibrations, offering a theoretical foundation and pragmatic recommendations for addressing vibrational challenges in real-world bridge engineering contexts.

\section{Assumptions And Notations in This Study}

This research employs the Euler-Bernoulli beam theory as its foundation, integrating specific assumptions to effectively model the plane bending vibration of beams. The primary assumptions underpinning this analysis are outlined as follows:

\begin{itemize}
    \item \textbf{Focus on Plane Bending Vibration}: This analysis exclusively addresses the plane bending vibrations manifest within the beam's principal plane. It adheres to the principles of minimal deformation and plane assumptions, as delineated in the discipline of material mechanics.
    \item \textbf{Consistency in Sign Convention}: The designation of positive symbols is aligned with the conventions established in material mechanics.
    \item \textbf{Linear Elastic Behavior}: The beam is hypothesized to display linear elastic characteristics, denoting a directly proportional relationship between stress and strain.
    \item \textbf{Small Deflection Hypothesis}: It is posited that the beam's deflection, during bending vibration, is negligible compared to its original length.
    \item \textbf{Homogeneity in Material and Cross-Section}: The beam is presumed to maintain uniform material properties and geometric dimensions throughout its length, ensuring homogeneity.
    \item \textbf{Planar Cross-Section Postulate}: Following bending, each cross-section of the beam is assumed to retain its planar nature and remain perpendicular to the vertical axis.
    \item \textbf{Exclusion of External Damping Effects}: The model presumes the absence of any external damping phenomena, such as air resistance.
    \item \textbf{Z-Axis Displacement Consideration}: The displacement of any point along the beam's axis is contemplated solely in terms of its z-axis orientation.
\end{itemize}

The following table delineates the symbols utilized in this article, accompanied by their respective descriptions:

\begin{table}[h]
\centering
\begin{tabular}{cl}
\hline
Symbol & Description \\
\hline
\( M \) & Bending moment \\
\( Q \) & Shear force \\
\( u \) & Displacement of points on the axis \\
\( \omega \) & Natural frequency \\
\( EI \) & Bending stiffness \\
\( \rho \) & Density \\
\( A \) & Cross section area \\
\hline
\end{tabular}
\end{table}

\section{Model Building}
\subsection{Establishment of lateral vibration model for bridges}
\begin{figure}[ht!] 
\centering
\includegraphics[width=3.0in]{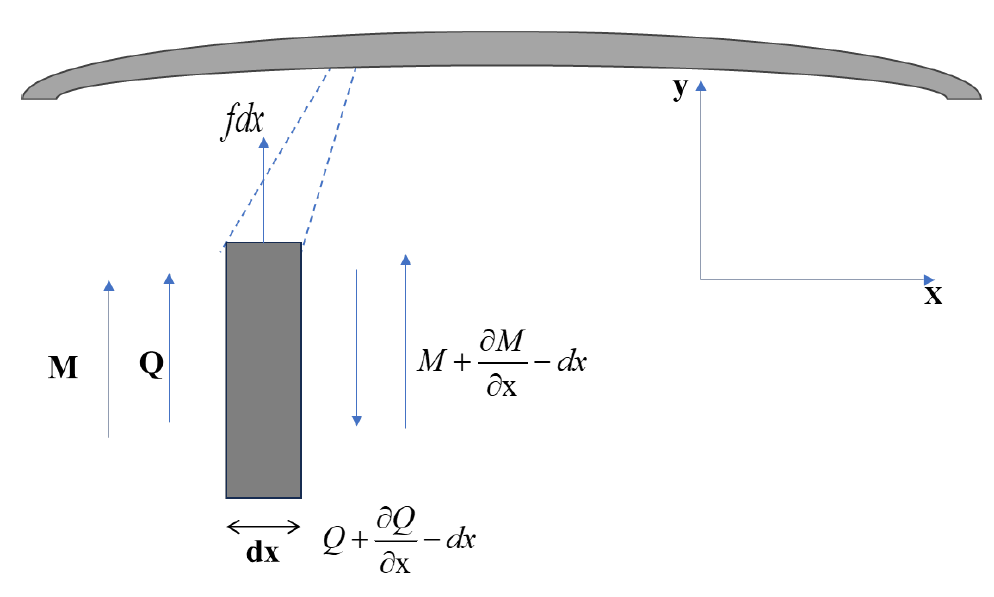}
\caption{Lateral Vibration Model For Bridges}
\label{1}
\end{figure}

As depicted in Figure \ref{1}, employing the differential element method enables the segmentation of the beam into an infinitesimal length \(dx\). Within this framework, \(M\) and \(Q\) are introduced, adhering to the sign conventions consistent with the previously stated assumptions. The motion equation for the beam segment \(dx\) in the z-direction is formulated as follows:
\begin{equation}
\frac{\partial Q}{\partial x} = -\rho A \frac{\partial^2 u}{\partial t^2} + f
\end{equation}
Then, using the relationship in material mechanics,
\begin{equation}
Q = \frac{\partial u}{\partial x}, M = EI\frac{\partial^2 u}{\partial x^2}
\end{equation}
The bending vibration equation of the beam is obtained as:
\begin{equation}
\frac{\partial^2}{\partial x^2}\left[EI\frac{\partial^2 u}{\partial x^2}\right] = -\rho A\frac{\partial^2 u}{\partial t^2} + f
\end{equation}

Resembling one-dimensional wave equations, the resolution of a definite solution necessitates specific boundary and initial conditions. Commonly encountered conditions include:

\begin{itemize}
    \item Simply supported end: The deflection and bending moment are 0, i.e.:
    \begin{equation}
    \mu{}(0,t) = 0, EI\frac{\partial^2 u(x, t)}{\partial x^2}\bigg|_{x=0} = 0
    \end{equation}
    
    \item Free end: The bending moment and shear force are 0, i.e.:
    \begin{equation}
    EI\frac{\partial^2 u(x_1t)}{\partial x^2}\bigg|_{x=0} = 0, \frac{\partial}{\partial x}\left[EI\frac{\partial^2 u(x_1t)}{\partial x^2}\right]\bigg|_{x=0} = 0
    \end{equation}
    
    \item Fixed end: The deflection and rotation angle are 0. That is:
    \begin{equation}
    \mu{}(0,t) = 0, \frac{\partial u(x, t)}{\partial x}\bigg|_{x=0} = 0
    \end{equation}
\end{itemize}

Initially, the analysis is simplified by considering the interference force in the vibration equation as negligible:
\begin{equation}
\frac{\partial^2}{\partial x^2} \left( EI \left[ \frac{\partial^2 u}{\partial x^2} \right] \right) = -\rho A \frac{\partial^2 u}{\partial t^2}
\end{equation}

Further simplification leads to the vibration equation:
\begin{equation}
a^2\frac{\partial^4 u}{\partial x^4} + \frac{\partial^2 u}{\partial t^2} = 0, \quad a = \sqrt{\frac{EI}{\rho A}}
\end{equation}

Application of the variable separation method results in:
\begin{equation}
u(x, t) = \phi(x) q(t)
\end{equation}

Subsequent simplification and substitution yield:
\begin{equation}
\frac{a^2 \frac{\partial^2}{\partial x^2} \left[ \frac{d^2 \phi (t)}{dx^2} \right]}{\phi (x)} = \frac{- \frac{d^2 q(t)}{dt^2}}{q (t)} = w^2
\end{equation}

This process culminates in two distinct equations:
\begin{equation}
\left\{
\begin{array}{l}
\frac{d^2 q(t)}{dt^2} + w^2 q(t) = 0 \\
\frac{d^4 \phi(x)}{dx^4} = \beta^4 \phi(x)
\end{array}
\right.
\end{equation}
where \( \beta^4 = \frac{w^2}{a^2} \).

The general solution is thus represented as:
\begin{equation}
\begin{split}
\phi(x) &= c_1 \sin \beta x + c_2 \cos \beta x + c_3 \sinh \beta x + c_4 \cosh \beta x \\
q(t) &= c_5 \sin wt + c_6 \cos wt
\end{split}
\end{equation}

Afterwards, consider the boundary conditions and the definite solution situation. 
\begin{figure}[ht!] 
\centering
\includegraphics[width=3.0in]{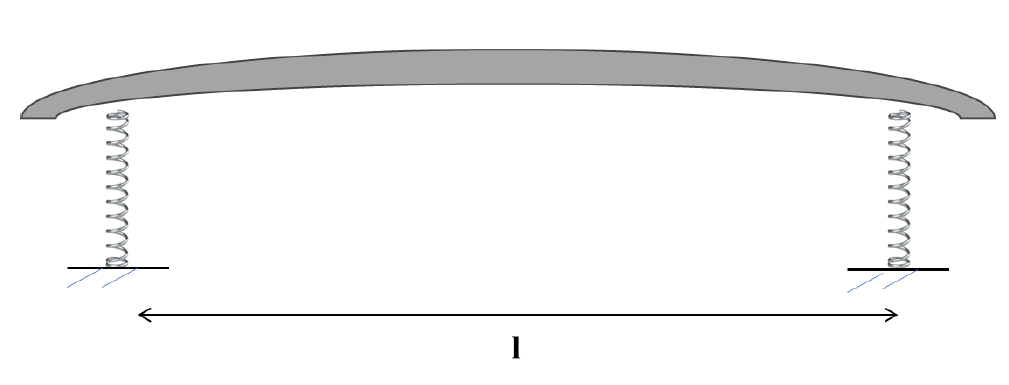}
\caption{Boundary Condition Analysis of a Supported Beam}
\label{2}
\end{figure}
Figure \ref{2} illustrates the boundary condition analysis of a supported beam. The beam is analyzed under conditions where bending moments at both ends are null and shear forces equal the elastic forces. This scenario exclusively considers the case where both segments are bearings, with other boundary conditions derivable in a similar fashion. (Note: Here, only four equations are present, hence a constant \( c \) is used for reference. In this article, \( c_2 \) represents \( c_1, c_3 \), and \( c_4 \)).

The boundary conditions at both ends are established as follows: the bending moment is null, and the shear force equals the elastic force.
\begin{equation}
\begin{aligned}
\phi''(0) &= 0, \quad Q = \frac{du}{dx}= \left[ EI q\frac{d^3 \phi}{dx^3} \right]_{x=0} = qk\phi{}(0) \\
\phi''(1) &= 0, \quad Q = \frac{du}{dx}= \left[ EI q\frac{d^3 \phi}{dx^3} \right]_{x=l} = qk\phi{}(l)
\end{aligned}
\end{equation}

Characteristic equation derivation:
\begin{equation}
\begin{aligned}
\phi'(x) &= C_1\beta \cos \beta x - C_2\beta \sin \beta x + C_4\beta \beta x + C_3\beta \beta x \\
\phi''(x) &= -C_1 \beta^2 \sin \beta x - C_2 \beta^2 \cos \beta x + C_3 \beta^2\beta{}x + C_4 \beta^2\beta{}x \\
\phi'''(x) &= -C_1 \beta^3 \cos \beta x + C_2 \beta^3 \sin \beta x + C_3 \beta^3\beta{}x + C_4 \beta^3\beta{}x
\end{aligned}
\end{equation}

Substituting the boundary conditions results in the following equation system:
\begin{equation}
\begin{aligned}
\begin{pmatrix}
0 & -\beta^2 & 0 & \beta^3 \\
-\beta^2 \sin(\beta l) & -\beta^2 \cos(\beta l) & \beta^2 l & \beta^2 l \\
-EI\beta^3 & 0 & EI\beta^3 & 0 \\
-EI\beta^3 \cos(\beta l) & EI\beta^3 \sin(\beta l) & EI\beta^3 l & EI\beta^3 l
\end{pmatrix}
\begin{pmatrix}
c_1 \\
c_2 \\
c_3 \\
c_4
\end{pmatrix}
\\=k
\begin{pmatrix}
0 \\
0 \\
 \begin{vmatrix} 1 & 1 \\ c_2 & c_4 \end{vmatrix} \\
 \begin{vmatrix}
\sin(\beta l) & \cos(\beta l) & \sinh(\beta l) & \cosh(\beta l) \\
c_1 & c_2 & c_3 & c_4
\end{vmatrix}
\end{pmatrix}
\end{aligned}
\end{equation}

The structural properties of the given equation facilitate a process of further simplification, leading to a refined set of relations as follows:
\begin{equation}
\left\{
\begin{aligned}
& c_2 = c_4 \\
& c_2 \beta I - c_2 \cos(\beta I) + c_3 \sinh(\beta I) - c_1 \sin(\beta I) = 0 \\
& M=c_1 \sin(\beta I) + c_2 \cos(\beta I) + c_3 \sinh(\beta I) + c_4 \cosh(\beta I)\\
& \frac{-c_1 + c_3}{M} = \frac{c_2 + c_4}{M}
\end{aligned}
\right.
\end{equation}

In this context, let us consider $c_2$ as a representative constant encompassing various quantities. Consequently, the expression:
\begin{equation}
c_3=\cfrac{c_1sin\beta{}l+_c2cos\beta{}l-c_2ch\beta{}l}{sin\beta{}l}
\end{equation}
can be substituted into the above fraction to obtain the following equation:
\begin{equation}
\left\{
\begin{aligned}
& A=\cfrac{c_2ch\beta{}l-c_2cos\beta{}l+c_3sh\beta{}l}{sin\beta{}l} \\
& B=c_2sin\beta{}l+c_3sin\beta{}l+c_2sh\beta{}l \\
& C=c_2cos\beta{}l+c_3sh\beta{}l+c_2ch\beta{}l \\
& -2c_2A+2c_2B=-Asin\beta{}l-A(C+c_3sin\beta{}l)+c_3C
\end{aligned}
\right.
\end{equation}

The task at hand involves solving this system of quadratic equations, and subsequently denoting $c_3$ with $c_2$, while also representing $c_1$ and $c_4$.

The parameter $\beta$ can be deduced from the aforementioned results. By applying the principle of fraction nature, the variables $c_1$, $c_2$, $c_3$, and $c_4$ can be systematically eliminated.

This leads to the substitution into the expression for calculating frequency:
\begin{equation}
\omega{}=\beta{}^{2}a=\beta{}^{2}\sqrt{\cfrac{EI}{\rho{}A}}
\end{equation}

From the above derivation, it becomes evident that when dealing with a plethora of parameters and a high number of partial differential equations, traditional computational methods for solving definite solution problems face considerable challenges, particularly when confronting complex quadratic or even higher-order equations. To address this complexity, this article proposes a solution that leverages numerical simulation formulas, accompanied by further condition constraints to streamline the process.

\subsection{Resonance Principle}
Bridge resonance is a multifaceted issue, encompassing aspects of structural dynamics and vibrational mechanics. Fundamentally, when a bridge is subjected to periodic loads that align with its natural frequency, it can enter a state known as resonance, as delineated by Zhou (2023) in their study on structural damage\cite{zhou2023damage}. In this state, the vibrational amplitude of the bridge escalates markedly. This escalation can precipitate structural impairment or, in severe cases, total collapse, particularly in scenarios where damping mechanisms are inadequate to mitigate the vibrations.

The core principles of resonance can be distilled into the following key concepts:
\begin{itemize}
    \item \textbf{Natural frequency}:This refers to the inherent vibrational frequency of a structure that occurs in the absence of external forces. In the context of bridges, this frequency is predominantly influenced by the bridge’s geometric configuration, material properties, and support conditions.
    \item \textbf{Resonance}: This phenomenon arises when an external periodic load—emanating from sources such as vehicular traffic, wind forces, or seismic activity—matches or closely approximates the natural frequency of the structure. During resonance, the amplitude of structural vibrations can increase substantially, potentially leading to significant damage or catastrophic failure.
\end{itemize}

To elucidate the principle of resonance more comprehensively, one could consider the application of a periodic force to a structural model, observing how it interacts with the structure's natural vibrational characteristics. This approach would provide an insightful demonstration of the conditions under which resonance occurs and its impact on structural integrity.

\subsection{Establishment of Resonance Phenomenon}
Given the complexity of the definitive solution to the equation previously established, complemented by the response force, an alternative approach is proposed to elucidate the phenomenon of resonance \cite{wei2023modification}.

The discourse herein utilizes a simplified single degree of freedom (SDOF) model to approximate the dynamic response of bridge structures. This model encapsulates the dynamics of a point mass system characterized by mass \( m \), stiffness \( k \), and damping \( c \). The governing equation of motion is articulated as follows:
\begin{equation}
m\ddot{u}(t) + c\dot{u}(t) + ku(t) = F(t)
\end{equation}

In practice, however, the dynamic representation of bridges often necessitates the consideration of multidirectional vibrations and motions, encompassing multiple degrees of freedom. This includes but is not limited to lateral, longitudinal, and torsional modes of motion. To accurately represent such multidimensional dynamics, this study extends to a multi-degree of freedom (MDOF) system, building upon the foundational SDOF model.

The MDOF framework disaggregates the structural system into an assembly of discrete mass particles interconnected by elastic (spring) and dissipative (damping) elements. Each particle is afforded the capability to exhibit motion in multiple vectors, with each vector representing a singular degree of freedom. The dynamic equations governing this system, considering a bidirectional degree of freedom for illustrative purposes, are expressed as:
\begin{equation}
\begin{bmatrix}
m & 0 \\
0 & m 
\end{bmatrix}
\begin{bmatrix}
\ddot{x} \\
\ddot{y}
\end{bmatrix}
+
\begin{bmatrix}
c & 0 \\
0 & c 
\end{bmatrix}
\begin{bmatrix}
\dot{x} \\
\dot{y}
\end{bmatrix}
+
\begin{bmatrix}
k & 0 \\
0 & k 
\end{bmatrix}
\begin{bmatrix}
x \\
y
\end{bmatrix}
=
\begin{bmatrix}
F_x(t) \\
F_y(t)
\end{bmatrix}
\end{equation}

\subsection{Methods in Numerical Simulation}

Numerical simulation methodologies play a pivotal role in the analysis of structural dynamics and response. Among these, the Finite Element Method (FEM) is preeminent for spatial discretization, segmenting a structure into a collection of finite elements and enabling the construction of a global system equation through the assembly of individual element equations. Modal analysis offers a reductionist approach, facilitating the simplification of a Multi-Degree of Freedom (MDOF) system to a Single-Degree of Freedom (SDOF) model, thus streamlining the analytical process. Direct time integration methods, exemplified by the \textit{Newmark} \( \beta \) Method, are instrumental in computing the responses of MDOF systems.

With consideration of various analytical perspectives, the \textit{Newmark} \( \beta \) method is employed herein.\cite{pourzeynali2021comprehensive} The essence of this method lies in its capability for direct time integration of dynamic equations, aptly handled with a discrete time increment \( \Delta t \). The method's operational formulae, which estimate the system's state at \( t + \Delta t \), are encapsulated as follows:

\begin{equation}
\left\{
\begin{aligned}
& u(t + \Delta t) = u(t) + \Delta t \cdot \dot{u}(t) + \frac{\Delta t^2}{2} \mathbf{a}_{\beta}, \\
& \dot{u}(t + \Delta t) = \dot{u}(t) + \Delta t \cdot \mathbf{a}_{\gamma}, \\
& \mathbf{a}_{\beta} = (1 - 2\beta)\ddot{u}(t) + 2\beta \cdot \ddot{u}(t + \Delta t), \\
& \mathbf{a}_{\gamma} = (1 - \gamma)\ddot{u}(t) + \gamma \cdot \ddot{u}(t + \Delta t).
\end{aligned}
\right.
\end{equation}

\begin{algorithm}
\caption{Numerical Simulation Process Pseudocode}
\begin{algorithmic}[1]
\State Establish initial conditions encompassing displacement, velocity, acceleration, and physical parameters such as mass, stiffness, and damping.
\State Define the external forces acting on the structure.
\State Apply the Finite Element Method to discretize the bridge structure into finite elements.
\State Construct the global matrices representing mass, stiffness, and damping.
\State Initialize vectors for displacement, velocity, and acceleration.
\For{each time increment}
    \State a. Update the displacement, velocity, and acceleration using the \textit{Newmark} \( \beta \) method.
    \State b. Document or output the resultant vectors.
\EndFor
\State Synthesize or interpret the simulation results.
\end{algorithmic}
\end{algorithm}

\section{Numerical Experimental Simulation}
\subsection{Experiment 1: Lateral Load Response Analysis}

The foundation of this experiment is outlined by specific parameters, as delineated in Table \ref{tab:experiment1}. The deflection \( w(x, t) \) will be observed within the spatial range of \( 0 \leq x \leq L \) and within the time range of \( t \in [0, 10]\,\text{s} \). The initial condition is a stationary beam with \( w(x, 0)=0 \) and \( \frac{\partial w(x,0)}{\partial t} = 0 \).

\begin{table}[h]
\centering
\caption{Parameters for Experiment 1}
\label{tab:experiment1}
\begin{tabular}{l l}
\hline
\textbf{Parameter} & \textbf{Value} \\
\hline
Beam Length (\( L \)) & \( 10\,\text{m} \) \\
Distributed Load (\( q \)) & \( 5000\,\text{N/m} \) \\
Elastic Modulus (\( E \)) & \( 25 \times 10^9\,\text{Pa} \) \\
Section Width (\( b \)) & \( 0.2\,\text{m} \) \\
Section Height (\( h \)) & \( 0.4\,\text{m} \) \\
Elastic Bearing Stiffness (\( k \)) & \( 1000\,\text{N/m} \) \\
Density (\( \rho \)) & \( 2500\,\text{kg/m}^3 \) \\
\hline
\end{tabular}
\end{table}

The methodology will include:

\begin{itemize}
    \item \textbf{Discretization of Space and Time:} Division of the spatial and temporal ranges into small intervals for precise deflection calculation.
    \item \textbf{Initialization:} Setting the initial deflection and velocity of the beam to zero at \( t=0 \).
    \item \textbf{Iteration:} Application of the finite difference method for successive updating of the beam's deflection and velocity at each time step.
    \item \textbf{Observation and Analysis:} Monitoring of the deflection distribution at various time points and analysis of the beam's dynamic behavior.
\end{itemize}

\begin{figure}[ht]
    \centering
    \includegraphics[width=0.45\textwidth]{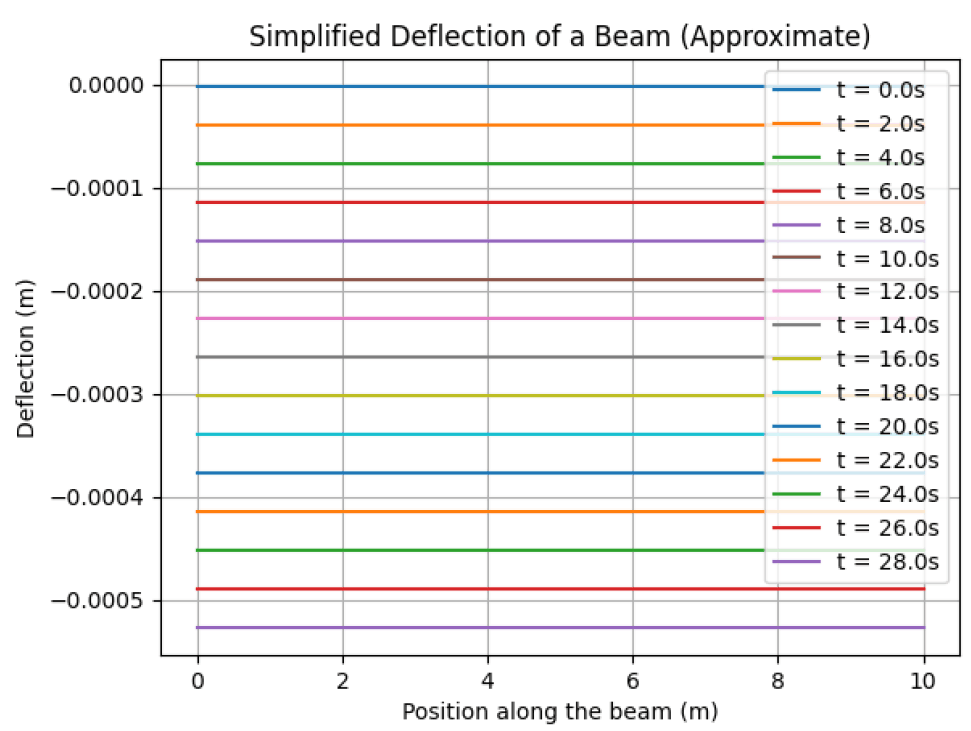}
    \caption{Simplified Deflection of a Beam (Approximate)}
    \label{3}
\end{figure}

The results of Experiment 1 are depicted in Figure \ref{3}. The graph demonstrates a consistent deepening in the beam's deflection over time under a constant load, illustrating a stable dynamic response. The deflection trends downwards, signifying the influence of the applied load, with the midspan likely exhibiting peak deflection. This underscores the efficacy of the finite difference method in predicting the behavior of a beam under specified conditions.

\subsection{Experiment 2: Beam Dynamics Under Moving Loads}
\subsubsection{Methodological Framework for Dynamic Load Simulations}

In the realm of practical engineering, bridge structures frequently encounter dynamic loads, primarily originating from vehicular traffic. This study rigorously investigates the effects of a moving concentrated load on a beam, simulating the dynamic loading conditions encountered in bridges with enhanced precision. The subsequent sections delineate the sophisticated methodologies and computational techniques employed to model and analyze this dynamic scenario.

\textbf{Defining Load Characteristics:} The success of this simulation hinges on accurately defining the parameters of the moving load. These parameters are detailed in Table \ref{tab2}:

\begin{table}[ht]
\centering
\caption{Definition of Load Characteristics}
\label{tab2}
\begin{tabular}{l l}
\hline
\textbf{Parameter} & \textbf{Meaning} \\
\hline
Load Magnitude (\( P \)) & Magnitude of the load \\
Load Velocity (\( v \)) & Velocity of the load’s traversal \\
Initial Load Position (\( x_0 \)) & Initial position of the load \\
\hline
\end{tabular}
\end{table}

These parameters are pivotal in constructing a realistic model that accurately reflects the dynamics of the load.

\textbf{Updating Load Position:} Essential to the simulation is the continuous updating of the load's position at each discrete time increment. This is governed by the equation:
\begin{equation}
x_p(t) = x_0 + v \times t
\end{equation}
where \( x_p(t) \) denotes the load's position at a given time \( t \), dynamically illustrating the load’s trajectory over time.

\textbf{Computational Response Analysis:} Capturing the dynamic impact of the moving load necessitates computing the structural response at each time step. This involves solving a time-dependent dynamic equation, enabling a thorough understanding of the structure's behavior under various load conditions.

\textbf{Interpretation of Structural Response:} The simulation's results, particularly the beam's deflection profile at different time intervals, are meticulously examined. This analysis concentrates on key performance metrics such as maximum deflection and the rate of deflection variation, providing insights into the structural response to dynamic loading.

\subsubsection{Experiment 2.1. Dynamic Deflection Analysis under Moving Load}

This experiment investigates the deflection response of a simply supported beam subjected to a single-point load traveling uniformly across it. The focus is on understanding how the dynamic load influences the structural deformation over time.

\textbf{Assumptions and Parameters:} The experimental model rests on specific assumptions, with parameters detailed in Table \ref{tab3}.

\begin{table}[ht]
\centering
\caption{Assumptions and Parameters for Beam Deflection Analysis}
\label{tab3}
\begin{tabular}{l l}
\hline
\textbf{Parameter} & \textbf{Value} \\
\hline
Load Magnitude (\( P \)) & \( 10,000\,\text{N} \) \\
Load Velocity (\( v \)) & \( 1\,\text{m/s} \) \\
Initial Load Position (\( x_0 \)) & \( 0\,\text{m} \) \\
Beam Length (\( L \)) & \( 10\,\text{m} \) \\
Distributed Load (\( q \)) & \( 0 \) (considering only dynamic loads) \\
Observation Duration (\( T \)) & \( 15\,\text{s} \) \\
\hline
\end{tabular}
\end{table}

The beam is modeled as simply supported, precluding vertical displacement or bending moments at the boundaries. The dynamic response is approximated by evaluating the static reaction to the load's instantaneous position at each time increment.

\textbf{Methodology:} The investigative steps are as follows:
\begin{itemize}
    \item Load Position Update: The load's position is systematically updated in accordance with the time elapsed.
    \item Response Computation: The static response of the load is calculated at each time step.
    \item Resultant Analysis: The deflection pattern over time is graphically represented and critically examined.
\end{itemize}

\begin{figure}[ht]
    \centering
    \includegraphics[width=0.45\textwidth]{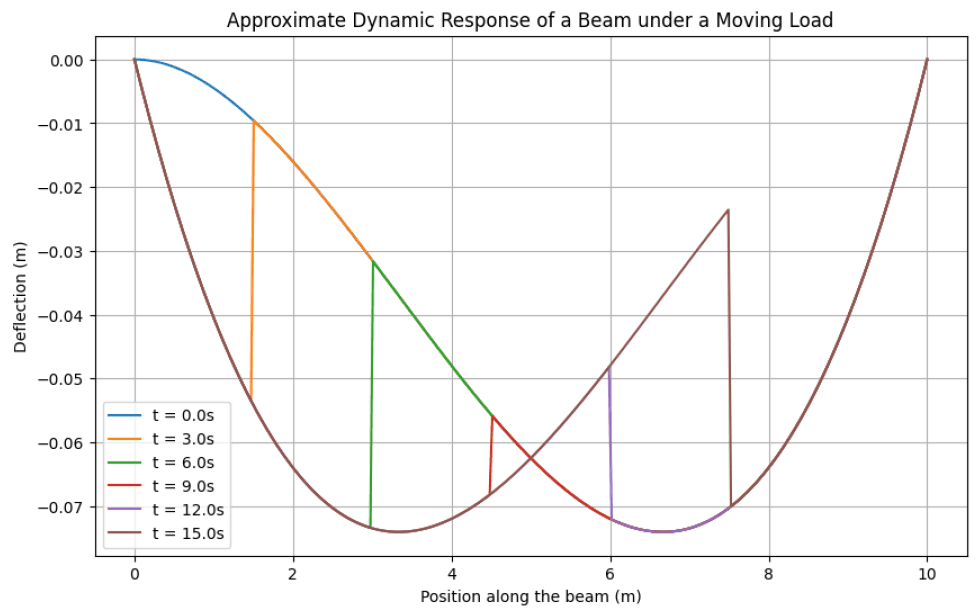}
    \caption{Approximate Dynamic Response of a Beam under a Moving Load}
    \label{fig:beam_deflection}
\end{figure}

As depicted in Figure \ref{fig:beam_deflection}, the beam exhibits a progressive deflection with the passage of the load, showing a pronounced deformation at the mid-span when the load is centrally located. The deflection curves at different time stamps elucidate the dynamic nature of the load's impact, with the maximum deflection occurring slightly after the load passes the beam's center due to the inherent dynamic effects and inertia of the structure.

\subsubsection{Experiment2.2. Sinusoidal Load Impact Analysis on Beam Deflection}
This segment of the study explores the dynamic response of a beam to a sinusoidal load, simulating the effects of periodic dynamic forces commonly seen in vibrating machinery and seismic activities on structural elements.

\textbf{Assumptions and Parameters:}The simulation is based on a set of predetermined assumptions and parameters encapsulated in Table \ref{tab4}.

\begin{table}[ht]
\centering
\caption{Assumptions and Parameters in Dynamic Load Simulation}
\label{tab4}
\begin{tabular}{l l}
\hline
\textbf{Parameter} & \textbf{Value} \\
\hline
Load Frequency \( f \) & 1 Hz \\
Load Amplitude \( P_0 \) & 10000 N \\
Load Position \( x \) & \( \frac{L}{2} \) m \\
Observation Time \( T \) & 10 seconds \\
Initial Condition & Stationary beam \\
\hline
\end{tabular}
\end{table}

\textbf{Methodology:}The load's magnitude at each time step is computed, subsequently allowing for the determination of the beam's deflection at that instance. The simulation employs a static response function to simplify the analysis.

\begin{figure}[h]
\centering
\includegraphics[width=0.88\linewidth]{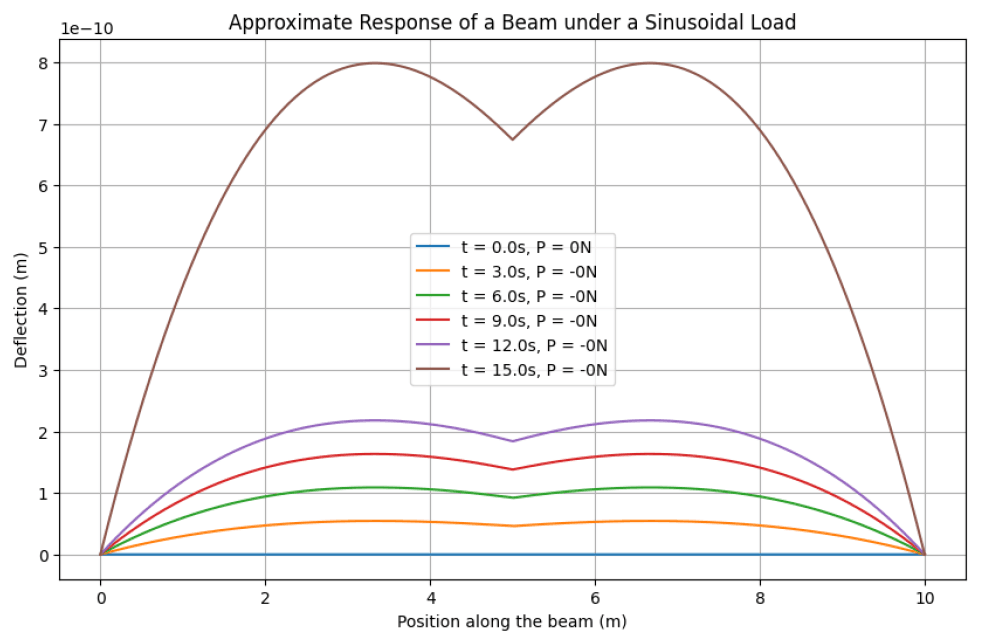}
\caption{Approximate Response of a Beam under a Sinusoidal Load}
\label{fig:sinusoidalLoad}
\end{figure}

As Illustrated in Figure \ref{fig:sinusoidalLoad}, the beam's deflection profiles exhibit a distinct pattern correlating with the sinusoidal nature of the applied load. At initial time increments, the deflection mirrors the load's amplitude, with the peak displacement occurring at mid-span, where the load is applied. As the simulation progresses, the response demonstrates a phase lag, attributed to the beam's inertia and damping characteristics. The deflection amplitude decreases over time, indicating the energy dissipation through the system and suggesting a stabilizing response of the beam to the cyclic loading.

\subsection{Experiment 3: Cantilever Beam Deflection under Concentrated Loads}
This experiment explores the deflection behavior of a cantilever beam when subjected to point loads. A cantilever beam, with one end fixed and the other free, provides a classic study in structural analysis. This research utilizes static analytical solutions to examine how varying the position of a concentrated load affects the beam's deflection, while simplifying the problem by excluding dynamic or distributed load effects.

The conditions and parameters underpinning this simulation are encapsulated in Table \ref{tab5}.

\begin{table}[ht]
\centering
\caption{Assumptions and Parameters for Cantilever Beam Analysis}
\label{tab5}
\begin{tabular}{l l}
\hline
\textbf{Parameter} & \textbf{Value} \\
\hline
Load Magnitude \( P \) & 10000 N \\
Load Position \( x \) & \( \frac{L}{2} \) m \\
Beam Length \( L \) & 10 m \\
\hline
\end{tabular}
\end{table}

\begin{figure}[ht]
\centering
\includegraphics[width=0.88\linewidth]{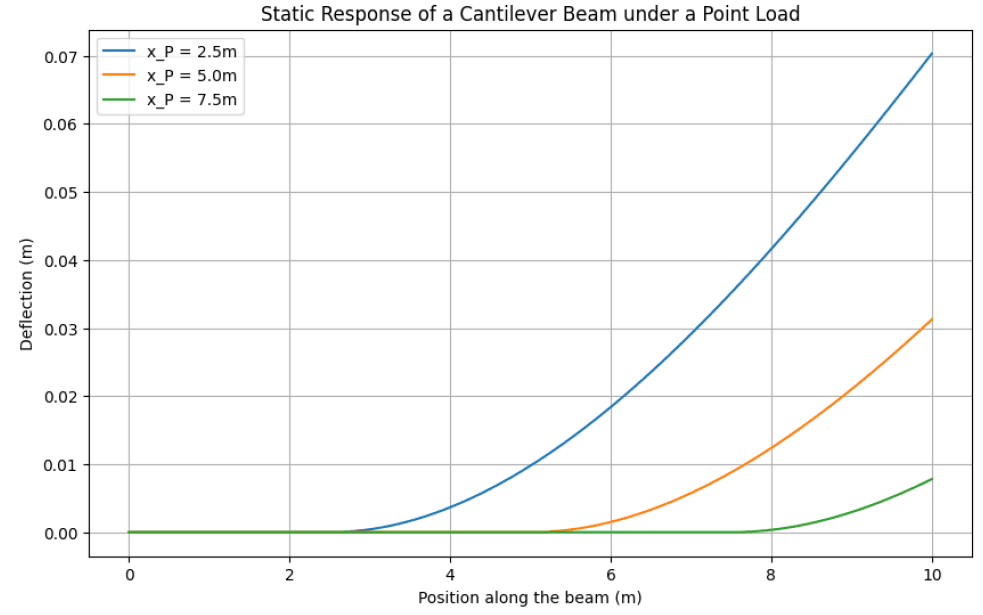}
\caption{Static Response of a Cantilever Beam under a Point Load}
\label{fig:cantilever_deflection}
\end{figure}

The results, depicted in Figure \ref{fig:cantilever_deflection}, show that the beam's deflection is significantly influenced by the load position. Deflections are measured at three distinct positions—near the fixed end, at mid-span, and close to the free end. The largest deflection occurs when the load is applied at the farthest point from the fixed support, highlighting the inverse relationship between deflection and distance from support. This finding is consistent with theoretical predictions and underscores the importance of load positioning in structural design and analysis.

\subsection{Experiment 4: Nonlinear Deflection Behavior of Cantilever Beams}
This experiment delves into the nonlinear deflection behavior of cantilever beams under point loads, contrasting the classical linear elastic model with a nonlinear stress-strain relationship. The study utilizes the Ramberg-Osgood model to describe the material's elastic-plastic transition, providing insight into the material's behavior beyond the proportional limit.

\begin{table}[ht]
\centering
\caption{Fundamental Assumptions and Parameters}
\label{tab:experiment4params}
\begin{tabular}{l l}
\hline
\textbf{Parameter} & \textbf{Value} \\
\hline
Beam Length \( L \)         & 10 m                    \\
Load Magnitude \( P \)      & 10000 N                 \\
Load Position \( x \)       & \( \frac{L}{2} \) m     \\
\hline
\end{tabular}
\end{table}

The Ramberg-Osgood model equation, \(\sigma = E\epsilon + \alpha E (\epsilon^{n})\), elucidates the nonlinearity in the material response, where \(\sigma\) represents stress, \(\epsilon\) strain, \(E\) the initial modulus of elasticity, and \(\alpha\) and \(n\) are parameters indicating material hardening.

\begin{figure}[ht]
\centering
\includegraphics[width=\linewidth]{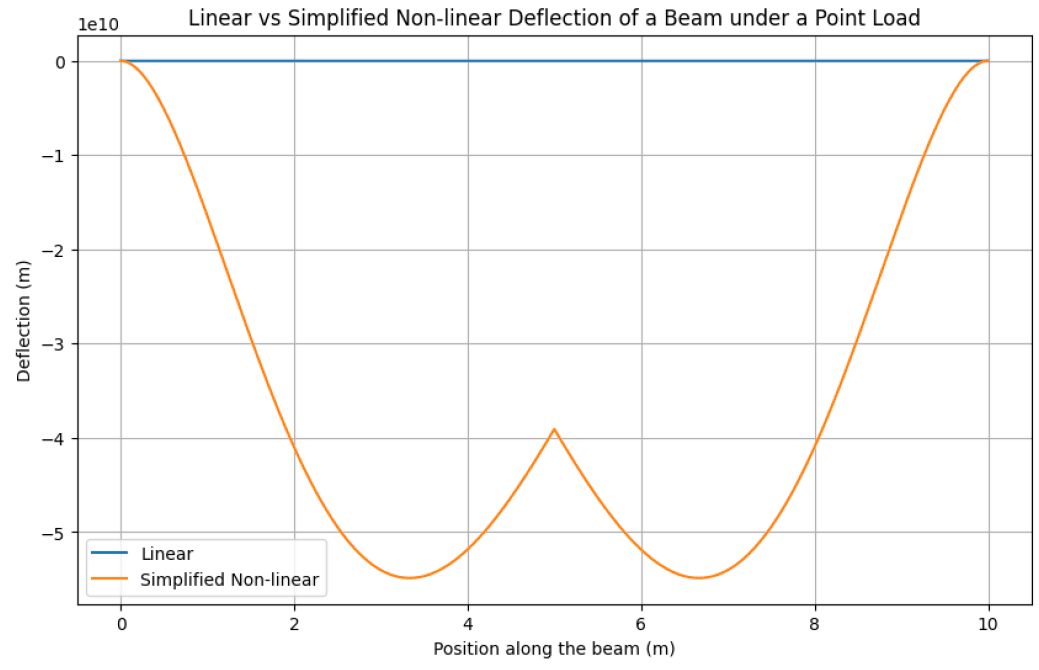}
\caption{Linear vs Simplified Non-linear Deflection of a Beam under a Point Load}
\label{fig:nonlinear_deflection}
\end{figure}

Figure \ref{fig:nonlinear_deflection} presents the comparative deflection profiles of the beam, revealing a marked deviation from linearity as the applied load increases. The linear model predicts a uniform increase in deflection, but the simplified nonlinear model shows a significant increase in deflection after a certain point, suggesting a substantial change in stiffness due to the material yielding. This behavior highlights the importance of considering nonlinear material properties for accurate structural analysis and design, especially for loads approaching or surpassing the yield strength.

\subsection{Experiment 5: Vibration Analysis of Beams and Structural Models}

\subsubsection{Experiment 5.1. Resonance Characterization in Simply Supported Beams}
This study examines the resonance behavior of a simply supported beam when subjected to sinusoidal loading at varying frequencies. The primary objective is to identify the beam's natural frequency using theoretical expressions and to observe the resonance phenomena that occur when the loading frequency approaches this natural frequency.

\begin{table}[ht]
\centering
\caption{Parameters for Resonance Analysis}
\label{tab:experiment5params}
\begin{tabular}{l l}
\hline
\textbf{Parameter} & \textbf{Value} \\
\hline
Beam Length \( L \) & 10 m \\
Elastic Modulus \( E \) & \( 25 \times 10^9 \) Pa \\
Section Width \( b \) & 0.2 m \\
Section Height \( h \) & 0.4 m \\
Material Density \( \rho \) & \( 2500 kg/m^3 \) \\
Load Amplitude \( P_0 \) & \( 1000 N \) \\
Load Frequency \( f \) & Varies \\
\hline
\end{tabular}
\end{table}

The resonance analysis covers a spectrum of frequencies, specifically targeting the first natural frequency of the beam as calculated by the given expression.

\begin{figure}[ht]
\centering
\includegraphics[width=\linewidth]{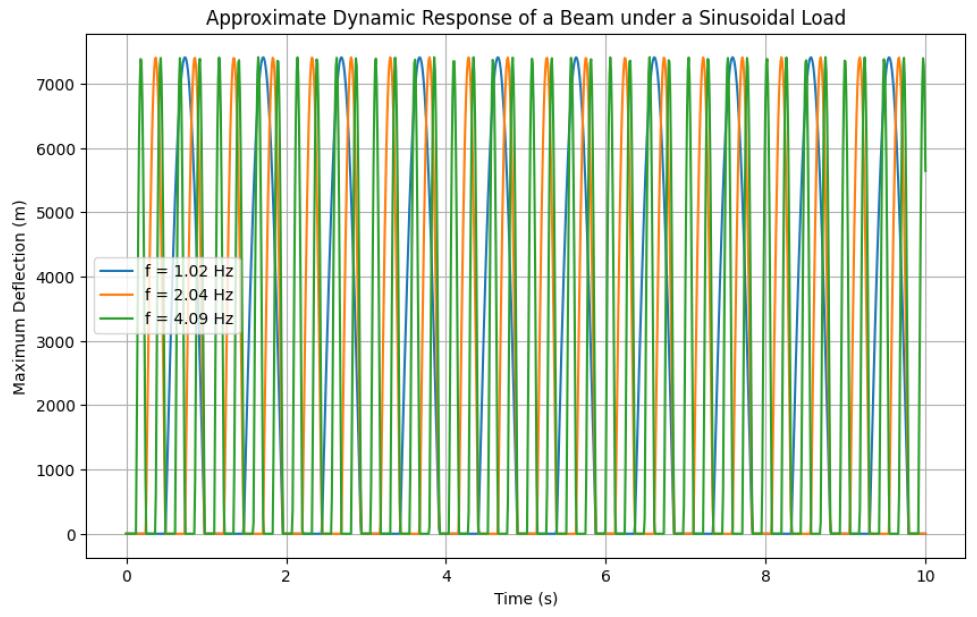}
\caption{Approximate Dynamic Response of a Beam under a Sinusoidal Load}
\label{fig:beam_resonance}
\end{figure}

As depicted in Figure \ref{fig:beam_resonance}, the beam's response to sinusoidal loading at different frequencies reveals a significant amplification in deflection at certain frequencies. The graph shows peak deflection amplitudes at frequencies of 1.02 Hz, 2.04 Hz, and 4.09 Hz, which correspond to the fundamental and higher modes of vibration. Notably, the maximum deflection is observed at the beam's first natural frequency, validating the theoretical prediction. This amplification at resonance underscores the critical importance of understanding dynamic loading in structural design to mitigate potential resonant vibration damage.

\subsubsection{Experiment 5.2. Comparative Dynamic Response Analysis}
This experiment provides a comparative analysis of the dynamic responses exhibited by single-degree-of-freedom (SDOF) systems and two-dimensional (2D) bridge models when subjected to sinusoidal forces. This analysis is crucial for understanding how simple and complex structural systems react to dynamic loading, a common occurrence in engineering practice.

\begin{figure}[ht]
\centering
\includegraphics[width=\linewidth]{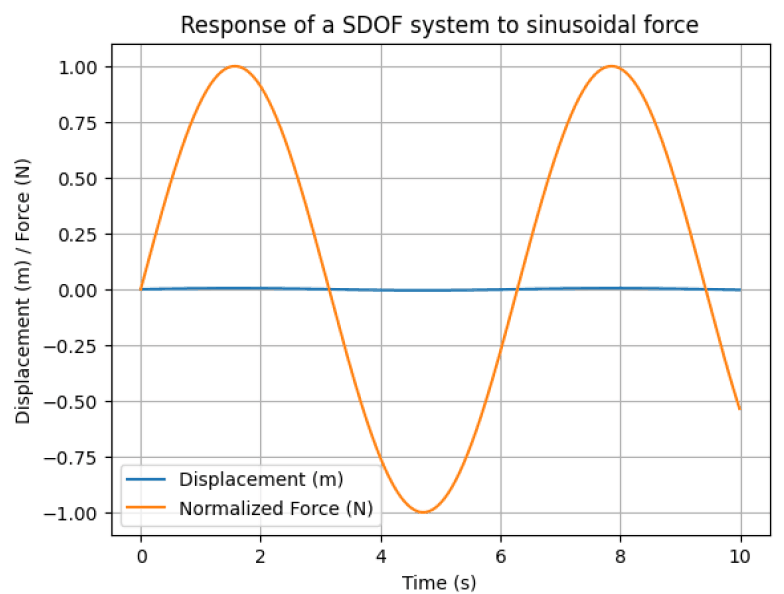}
\caption{Response of a SDOF system to sinusoidal force}
\label{fig:singleDOFResponse}
\end{figure}

\begin{figure}[ht]
\centering
\includegraphics[width=\linewidth]{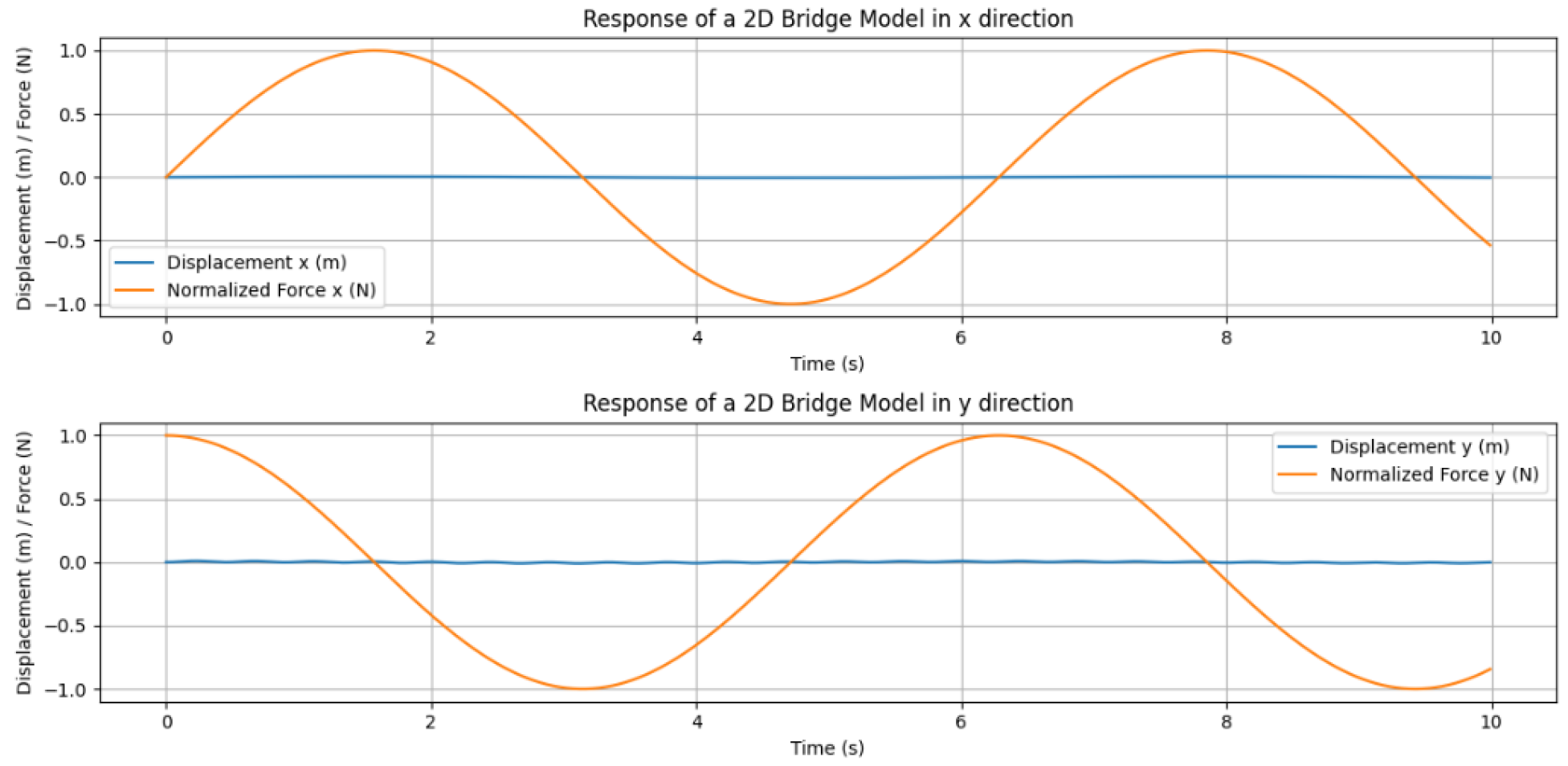}
\caption{Response of a 2D Bridge Model in x direction}
\label{fig:2DBridgeModelResponse}
\end{figure}

\textbf{Experimental Results:} The responses, as shown in Figures \ref{fig:singleDOFResponse} and \ref{fig:2DBridgeModelResponse}, highlight the contrasting behavior between the SDOF system and the 2D bridge model. For the SDOF system, there is a clear periodic displacement that mirrors the sinusoidal input force, indicative of a resonant frequency phenomenon. In contrast, the 2D bridge model demonstrates a more complex response, with the displacement in the x-direction showing a phase shift and amplitude variation, implying the presence of multiple modes of vibration and damping effects. These observations underscore the necessity for sophisticated analysis methods to predict the behavior of complex structures accurately.

The experiment substantiates the theoretical premise that a system's response to dynamic loading is profoundly influenced by its modal characteristics. The SDOF system, with its singular mode of vibration, presents a straightforward case study. Meanwhile, the 2D bridge model, embodying a more intricate dynamical system, provides insight into the nuanced responses of real-world structures, thus informing better design practices to mitigate potential resonance-related issues.

\section{Conclusion}
This study has methodically investigated the dynamic behavior of beams under diverse load conditions, revealing critical insights into the vibrational dynamics essential for civil engineering applications. Key experiments demonstrated the significance of accurately modeling dynamic responses to ensure the integrity of structures like bridges when subjected to vehicular traffic and other varying forces.

Experiments 1 and 2 showcased how deflections change with the speed and position of a moving load and the frequency of oscillating loads, establishing the need for precise analysis in structural design. Experiment 3's focus on cantilever beams under static loads offered a reminder of the necessity to incorporate a variety of external forces into design considerations for safety and stability.

In Experiment 4, the study of nonlinear material responses highlighted the deviations from expected linear behavior, especially beyond the elastic limit. The application of the Ramberg-Osgood model provided a reliable representation of such nonlinearity, underscoring the impact of material properties on structural responses.

Experiment 5's comparative analysis of SDOF and 2D bridge systems under sinusoidal forces emphasized the complexity of structural responses and the importance of modal analysis in anticipating and countering potential resonant-induced failures.

In essence, the collective findings from these experiments underpin the importance of a detailed and nuanced understanding of dynamic structural behavior, not just from a theoretical standpoint but as a practical imperative for designing robust structures. The insights from this research contribute towards the development of safer and more effective engineering designs, taking into account the dynamic nature of real-world loads.

This investigation reaffirms the synergy between theoretical models and empirical data as fundamental to the advancement of structural engineering, providing a basis for future innovations grounded in a solid understanding of dynamic responses.

\end{document}